\documentclass[twocolumn,amsmath,amssymb,prb]{revtex4-2}
\usepackage[utf8]{inputenc}
\usepackage{amsmath}
\usepackage{epsfig,amsmath,graphics}
\usepackage{epstopdf}
\usepackage{graphicx}
\usepackage{gensymb}
\usepackage{physics}
\usepackage{listings}
\usepackage{float}
\usepackage{enumitem}
\usepackage{color}
\usepackage{mathtools}

\begin{document}

\title{Multipartite Entanglement in the Random Ising Chain}

\author{Jay S. Zou}
\affiliation{Department of Physics and Astronomy, Northwestern University, Evanston, IL 60208}
\author{Helen S. Ansell}
\affiliation{Department of Physics and Astronomy, Northwestern University, Evanston, IL 60208}
\author{Istv\'an A. Kov\'acs}
\affiliation{Department of Physics and Astronomy, Northwestern University, Evanston, IL 60208}
\affiliation{Northwestern Institute on Complex Systems, Northwestern University, Evanston, IL 60208}

\date{\today}

\begin{abstract}
Quantifying entanglement of multiple subsystems is a challenging open problem in interacting quantum systems. 
Here, we focus on two subsystems of length $\ell$ separated by a distance $r=\alpha\ell$ and quantify their entanglement negativity (${\cal E}$) and mutual information (${\cal I}$) in critical random Ising chains. Both the disorder averaged ${\cal E}$ and ${\cal I}$ are found to be scale-invariant and universal, i.e. independent of the form of disorder. We find a constant ${\cal E}(\alpha)$ and ${\cal I}(\alpha)$ over any distances, using the asymptotically exact strong disorder renormalization group method. Our results are qualitatively different from both those in the clean Ising model and random spin chains of a singlet ground state, like the spin-$\frac{1}{2}$ random Heisenberg chain and the random XX chain. While for random singlet states ${\cal I}(\alpha)/{\cal E}(\alpha)=2$, in the random Ising chain this universal ratio is strongly $\alpha$-dependent. 
This deviation between systems contrasts with the behavior of the entanglement entropy of a single subsystem, for which the various random critical chains and clean models give the same qualitative behavior.
Therefore, studying multipartite entanglement provides additional universal information in random quantum systems, beyond what we can learn from a single subsystem.
\end{abstract}

\maketitle

\section{Introduction}

Entanglement is a distinguishing property of quantum mechanics, offering fundamentally stronger correlations than classical physics. In quantum many body systems, studying the entanglement properties is a promising way to understand universal properties, in particular the vicinity of quantum phase-transitions \cite{entanglement_review,amico,area,laflorencie}. In this paper, we show that quantifying multipartite entanglement provides additional universal insights, by distinguishing entanglement patterns that appear to be qualitatively the same when considering only a single subsystem.

Generally, the entanglement between a subsystem, ${A}$ and the rest of the system, ${ B}$,
in the ground state, $|\varPsi\rangle$ is quantified by the von Neumann entropy of the reduced density matrix,
$\rho_{A}={\rm Tr}_{{B}} | \varPsi \rangle \langle \varPsi |$ as
\begin{equation}
{\cal S}_A=-{\rm Tr}_{ A}\left(\rho_{A} \log_2{ \rho_{A}}\right)\;.
\label{eq:S}
\end{equation}
In one-dimensional systems we have an almost complete understanding \cite{holzhey,vidal,Calabrese_Cardy04}: ${\cal S}_A$ is known to diverge logarithmically at a quantum critical point as ${\cal S}_A=\frac{c}{3} \log_2 \ell+cst$. Here $\ell$ is the size of the subsystem and the prefactor is universal, $c$ being the central charge of the conformal field theory. These results have been extended to further properties, including the R\'enyi entropy and the properties of the entanglement spectrum \cite{Calabrese_Lefevre}.
%
%
Qualitatively similar results have been obtained for quantum models in the presence of quenched disorder \cite{refael}. In random chains (random antiferromagnetic Heisenberg and XX models,
random Ising model (RIM), etc.) the critical point is controlled by a so called
infinite disorder fixed point \cite{fisher, danielreview}, the properties of which can be conveniently
studied by the strong disorder renormalization group (SDRG) method \cite{mdh,im}.
Using the SDRG, a logarithmic entanglement entropy is found with a universal
prefactor \cite{refael_moore04}, which has been numerically checked by density-matrix
renormalization \cite{laflorencie} and by free-fermionic methods \cite{igloi_lin08}. It is not a coincidence that various random chains show similarities, as indicated by the exact relationship between the entanglement entropy of the random XX chain and the RIM \cite{connect}. Overall, quantum entanglement of a single interval is qualitatively similar in all these critical quantum systems. The question arises, whether this similarity extends to the entanglement properties of multiple intervals. In other words, do we gain qualitatively new insights about critical quantum systems by quantifying multipartite entanglement? Here, we show that the answer is positive: the multipartite entanglement structure is qualitatively different in these otherwise similar systems.

\begin{figure}[ht!]
\centering
\includegraphics[width=1\linewidth]{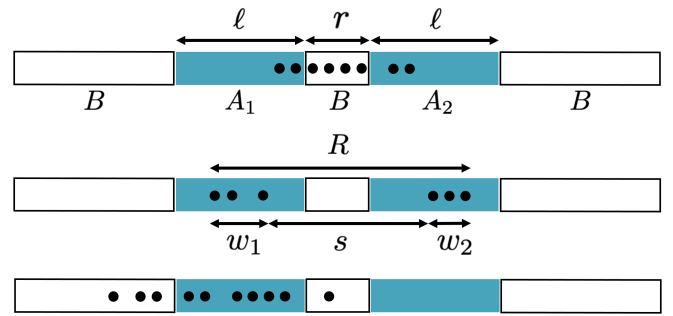}
\vskip-3mm
\caption{\label{fig:ill} We consider a tripartition of the system into two subsystems $A_1$ and $A_2$, both of length $\ell$, separated at a distance $r$, and the rest of the system, $B$. Top: illustration of a cluster in the RIM that only contributes to ${\cal I}$. Middle: a cluster that contributes to both ${\cal I}$ and ${\cal E}$. Bottom: a cluster that has no impact on either ${\cal E}$ or ${\cal I}$, see text for details.
}
\end{figure}

Quantifying multipartite entanglement is a challenging open task, even in small quantum systems \cite{szalay}.
Here, we consider two computable measures, the Entanglement Negativity (${\cal E}$) and Mutual Information (${\cal I}$) 
across multiple subsystems ($A_1$, $A_2$ and the rest of the system, $B$), see Fig.~\ref{fig:ill}. ${\cal I}$ is given by
\begin{equation}
    {\cal I}={\cal S}_{A_1}+{\cal S}_{A_2}-{\cal S}_{A_1\cup A_2},
        \label{eq:MI}
\end{equation}
while ${\cal E}$ is quantified by the logarithmic negativity 
\begin{equation}
    {\cal E}=\ln{\text{Tr}|\rho_{A_1 \cup A_2}^{T_2}|},
    \label{eq:EN}
\end{equation}
where $|\rho_{A_1 \cup A_2}^{T_2}|$ is the partially transposed reduced density matrix with respect to $A_2$. ${\cal E}$ is an entanglement monotone and it also serves as an upper bound for the distillable entanglement \cite{EN1,EN2,EN3,EN4,EN5,EN6}. In contrast, ${\cal I}$ accounts for both classical and quantum correlations, so a non-zero value does not necessarily mean quantum entanglement. 

Multipartite entanglement has been studied in random quantum chains that exhibit a so called \emph{random singlet} (RS) phase, where the ground state factorizes into singlets, each comprised of two spins, see for example the random XX-chain and random antiferromagnetic spin-$1/2$ Heisenberg (XXX) chain \cite{calabrese}.
In such random chains in the RS phase, ${\cal I}$ and ${\cal E}$ provide the same information as ${\cal I}/{\cal E}=2$, due to the simple entanglement structure, as shown by SDRG and free-fermionic techniques \cite{calabrese}. Here, we show that this is not true in general, as in the RIM ${\cal E}$ and ${\cal I}$ behave qualitatively differently. 

The RIM is given by the Hamiltonian
\begin{equation}
{\cal H} =
-\sum_{\langle ij \rangle} J_{ij}\sigma_i^x \sigma_{j}^x-\sum_{i} h_i \sigma_i^z\;,
\label{eq:H}
\end{equation}
in terms of the $\sigma_i^{x,z}$ Pauli-matrices at sites $i$ of a one-dimensional chain. The nearest neighbor couplings, $J_{ij}$, and the transverse fields, $h_i$,  are independent non-negative random numbers, taken from some non-singular distributions, to be specified later.  As far as universal properties are concerned, the shape of the distributions is irrelevant.
The ground state of the RIM is conveniently determined by an efficient SDRG algorithm \cite{alcaraz}. During the SDRG method \cite{im} the largest local terms in the Hamiltonian in Eq.~(\ref{eq:H}) are
successively eliminated and new Hamiltonians are generated through perturbation calculation. The critical properties of the RIM are governed by an infinite disorder fixed point, in which the strength of disorder grows without limit during renormalization \cite{danielreview}. Therefore, the SDRG results are 
asymptotically exact in the vicinity of the critical point, which is indeed demonstrated both analytically \cite{mccoywu,shankar} and numerically \cite{young_rieger96,bigpaper}. After decimating all degrees of freedom, the ground state of the RIM is found as a collection of independent ferromagnetic clusters of various sizes, each cluster being in a GHZ state $\frac{1}{\sqrt{2}}\left(|\uparrow \uparrow  \dots  \uparrow\rangle +
|\downarrow \downarrow \dots  \downarrow\rangle \right)$.

\section{Cluster Counting in the RIM}

For a single subsystem, each GHZ cluster contributes $\log_2 2=1$ to the entanglement entropy (Eq.~(\ref{eq:S})) if the cluster has at least one site inside and one site outside of the subsystem, otherwise the contribution is $0$ \cite{refael_moore04}. Thus, calculation of the entanglement entropy for the RIM is equivalent to a cluster counting problem. Here, we study ${\cal I}$ and ${\cal E}$ between two subsystems of size $\ell$, separated by distance $r$, with total system size $L$, shown in Fig.~\ref{fig:ill}. From Eq.~(\ref{eq:MI}) it readily follows that calculating ${\cal I}$ also leads to a cluster counting problem. Let's indicate the number of specific cluster configurations by $C_{x_1yx_2}$, where $x_i=1$ if the cluster has some sites in subsystem $A_i$, and $y=1$ if it has sites in $B$, while both $x_i$ and $y$ are 0 otherwise. Then, ${\cal S}_{A_1}=C_{110}+C_{101}+C_{111}$, ${\cal S}_{A_2}=C_{011}+C_{101}+C_{111}$, while ${\cal S}_{A_1\cup A_2}=C_{011}+C_{110}+C_{111}$, leading to 
\begin{equation}
{\cal I}=2C_{101}+C_{111}\;.  
\end{equation}

As ${\cal E}$ is additive on tensor products, independent magnetic clusters contribute additively. 
For ${\cal E}$, a non-vanishing contribution requires a cluster that has sites in both $A_1$ and $A_2$, while no sites outside \cite{calabrese}. Independently from the number of spins in the GHZ state, ${\cal E}$ is the same as for a Bell-state, which will have one negative eigenvalue of $-1/2$ after performing the partial transpose. 
Hence, ${\cal E}$ is the result of another cluster counting problem, ${\cal E}=C_{101}$, leading to
\begin{equation}
{\cal I}=2{\cal E}+C_{111}. 
\end{equation}
Interestingly, ${\cal I}-{\cal E}=C_{101}+C_{111}$ has the simplest geometric interpretation: the total number of clusters that have sites in both subsystems. Note that $C_{111}=0$ when clusters have only two sites, which is the case for the previously studied systems in the RS phase, leading to ${\cal I}=2{\cal E}$ \cite{calabrese}. In contrast, clusters in the RIM can have a large number of sites, yielding a non-zero $C_{111}$, breaking the simple proportionality of ${\cal I}$ and ${\cal E}$. At this point, it is not even clear if ${\cal I}$ or ${\cal E}$ (or their ratio) remain universal in the RIM.


\section{Analytic Results}
The SDRG has been used to provide the leading order scaling of the entanglement entropy for one subsystem \cite{refael}, corresponding to an effective central charge. 
Due to cluster counting, there is an underlying geometric interpretation 
\begin{equation}
    {\cal S}(L, \ell)
    =\frac{1}{L}\sum_{s=1}^{L/2}n(s)\min(\ell,s)\;,
    \label{eq:gap}
\end{equation}
where $n(s)$ is the number of instances in all clusters of `gap-size' $s$, defined as the distance between consecutive sites in the same cluster \cite{EPL}. With periodic boundary conditions this is given by $s=\min(x_{i+1}-x_i,L-x_{i+1}+x_i)$. 
For large $L$ the summation can be approximated by an integral in Eq.~(\ref{eq:gap}), yielding the leading order universal logarithmic scaling $S(L, \ell)=\frac{1}{6}\ln \ell$, as $n(s)=C s^{-2}$ for large $L$ and $C/L=c_{\text{eff}}/(3\ln2)=\frac{1}{6}$ for the RIM \cite{refael}. 

Based on the known gap-size statistics $n(s)$, a strict bound can also be calculated for ${\cal E}$ and ${\cal I}$. In practice, ${\cal E}$ and ${\cal I}$ are averaged over a large number of random samples as well as for all $L$ subsystem positions in each sample. First, we consider configurations where the subsystems are positioned in a way that the sites of the gap fall inside separate subsystems. This can happen only when $s>r$ in a cluster. 
Such a configuration has \emph{at least} $1$ contribution to ${\cal I}$ (2 if it has no sites in $B$, i.e. $C_{101}$), 
while \emph{at most} $1$ contribution to ${\cal E}$ (0 if it has sites in $B$, i.e. $C_{111}$). 
Out of all $L$ potential subsystem locations, there are $s-r$ (if $s<\ell+r$), or $2\ell+r-s$ (if $s\geq\ell+r$) such configurations. Hence, the bound is
\begin{align}
    \mathcal{B}=\frac{C}{L}\smashoperator{\int_{r}^{\ell+r}}\mathrm{d}s \frac{s-r}{s^2} + \frac{C}{L}\smashoperator{\int_{\ell+r}^{2\ell+r}}\mathrm{d}s \frac{2\ell+r-s}{s^2}\;.
\end{align}
$\mathcal{B}$ is a lower bound for ${\cal I}$ and an upper bound for ${\cal E}$. 
Performing the integrals and using $C/L=c_{\text{eff}}/(3\ln2)$, which is valid for both the critical RIM and critical RS states, yields
\begin{align}
    \mathcal{B}(x)=-\frac{c_{\text{eff}}}{3\ln2}\ln(1-x)\;.
\end{align}
Here $x$ is the scale-invariant cross ratio $x=\ell^2/(\ell+r)^2$, or in general
\begin{equation}
x=\frac{\ell_1\ell_2}{(\ell_1+r)(\ell_2+r)}, 
\label{eq:x}
\end{equation}
when the two subsystems have different lengths, $\ell_1$ and $\ell_2$. In an RS state, where each cluster has two sites, the bound captures ${\cal E}$ exactly for large $L$, $\mathcal{B}={\cal E}={\cal I}/2$, as shown independently in Ref.~\cite{calabrese}. 
In the case when $r=\alpha\ell$ and $\ell_1=\ell_2=\ell$, the bound is a distance-independent constant that only depends on $\alpha$ (Fig.~\ref{fig:alpha}):
\begin{align}
    \mathcal{B}(\alpha)=-\frac{c_{\text{eff}}}{3\ln2}\ln(1-\frac{1}{(1+\alpha)^2})\;.
\end{align}
Note that recently, in the RS phase, finite-size corrections to $n(s)$ have been characterized, also offering a way to calculate corrections to $\cal{B}$ \cite{corrections}.
In comparison, clusters in the RIM often contain more than two sites. As illustrated in Fig.~\ref{fig:ill}, a contributing gap is surrounded by a total linear extent of the cluster of length $w_1$ and $w_2$ on each side. Therefore, in the RIM the $\alpha$-dependence of ${\cal E}$ and ${\cal I}$ can be different from each other as well as that of $\cal{B}(\alpha)$, as discussed in more detail in the Appendix. 

\section{Numerical Results}
We have studied critical RIM chains up to subsystem size $\ell=8,192$ with at least $10,000$ realizations for each size, using periodic boundary conditions. Following Refs. \cite{2dRG,ddRG} we have used two ferromagnetic disorder distributions, where $J_{ij}$ is uniformly distributed in $\left[ 0,1\right]$. For \emph{box-$h$} disorder the distribution of the transverse-fields is uniform in $\left[ 0,h\right]$, whereas for the \emph{fixed-$h$} disorder we have $h_i=h,~\forall i$. The quantum control parameter is defined as $\theta=\ln(h)$, and the critical point is located at $\theta_{\text{box}}=0$ and $\theta_{\text{fixed}}=-1$, respectively. 

\begin{figure}[ht!]
\begin{center}
\includegraphics[width=1.0\linewidth]{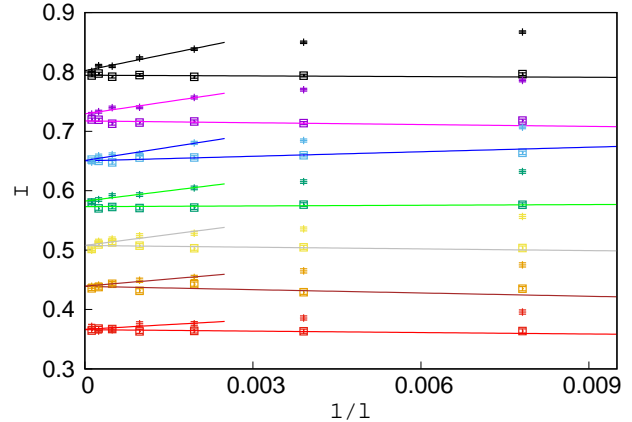}
\end{center}
\vskip-9mm
\caption{\label{fig:MI}
${\cal I}$ in the RIM for $\alpha=2^n$, with $n=-5,-4,\dots,1$ from top to bottom for fixed-$h$ ($+$) and box-$h$ ($\Box$) disorders.
}
\end{figure}

\begin{figure}[ht!]
\begin{center}
\includegraphics[width=1.0\linewidth]{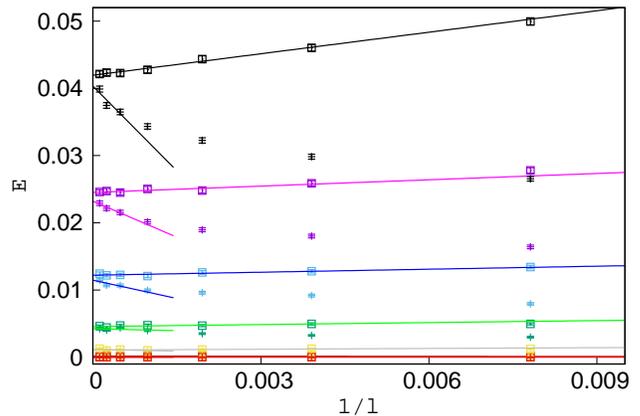}
\end{center}
\vskip-9mm
\caption{\label{fig:EN}
The same as Fig.~\ref{fig:MI} for ${\cal E}$. Interestingly, the finite-size corrections are of opposite sign compared to ${\cal I}$. For box-$h$ disorder, the linear extrapolation fits the data in the entire range of sizes for both ${\cal I}$ and ${\cal E}$. 
}
\end{figure}

\begin{figure}[ht!]
\begin{center}
\includegraphics[width=1.0\linewidth]{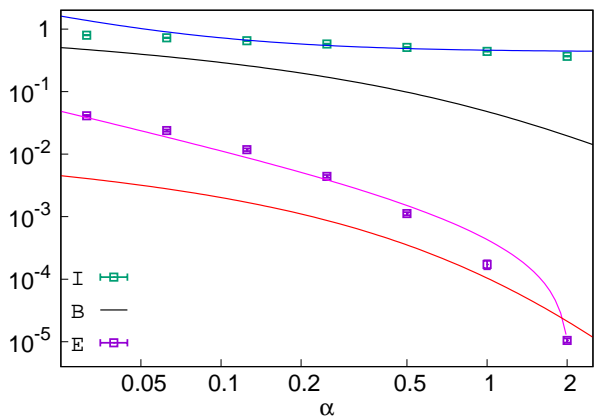}
\end{center}
\vskip-9mm
\caption{\label{fig:alpha}
The extrapolated ($1/\ell\to0$) ${\cal I}$ and ${\cal E}$ values, averaged over the two disorder realizations, as a function of $\alpha=r/\ell$, together with the analytic bound $\cal B$. As ${\cal I}\gg{\cal E}$, ${\cal I}$ can be fitted by Eq.~(\ref{MI-EN}) (blue line). ${\cal E}$ is fitted by two different approximations, given by Eq.~(\ref{eq:EN_app}) (magenta), as well as Eq.~(\ref{eq:EN_app2}) (red). In each case, the indicated error bars are smaller than the size of the symbol.
}
\end{figure}

Here, we present our results for the linear case, when $r=\alpha\ell$, with $\alpha=2^n$, sampled for $n=-5,-4,\dots,1$. The system size $L$ would ideally be much larger than $\ell$. Numerical SDRG studies for the entanglement entropy of a single subsystem of size $\ell$ have shown that $L=2\ell$ is sufficiently large in practice \cite{EPL}. Therefore, we chose $\ell$ such that the total linear extent of the boundaries of the two subsystems ($2\ell+r$) spans half of the system, i.e. $L=2(2\ell+r)$. Additional geometries are discussed in the next section. The finite-size estimates of ${\cal I}$ and ${\cal E}$ in the linear case are shown in Figs.~\ref{fig:MI} and \ref{fig:EN}, extrapolated to infinite size as $1/\ell\to0$. For both ${\cal I}$ and ${\cal E}$, the extrapolated values 
agree for the two disorder realizations, indicating universality. 
Our main results are summarized in Fig.~\ref{fig:alpha}, illustrating the universal ${\cal E}$ and ${\cal I}$ values for the studied values of $\alpha$. We observe that ${\cal I}$ and ${\cal E}$ are very different from each other, as well as from $\mathcal{B}$. In stark contrast to random quantum chains of a critical RS state, in the RIM the ratio of ${\cal I}$ and ${\cal E}$ depends on $\alpha$. Interestingly, in the studied range of $\alpha$, ${\cal I}$ appears to decay linearly as ${\cal I}=a-b\ln{\alpha}$, with $a=0.438(1)$ and $b=0.102(1)$. However, as ${\cal I}$ is positive and larger than $\mathcal{B}$, there must be deviations from this expression for larger values of $\alpha\sim50$. To support our numerical findings, an approximate argument for the $\alpha$-dependence is presented in the Appendix, providing Eq.~(\ref{MI-EN}), which was used in the fit in Fig.~\ref{fig:alpha}. The use of $\cal I-E$ as a proxy for $\cal I$ is justified by the fact that in the observed range ${\cal E}$ is much smaller than the upper bound provided by $\cal{B}$, see the Appendix for supporting analytic arguments. 

\section{Finite intervals}
\begin{figure}[ht!]
\begin{center}
\includegraphics[width=1.0\linewidth]{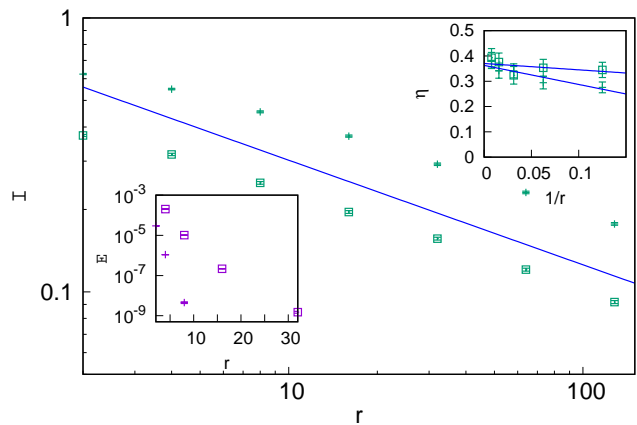}
\end{center}
\vskip-9mm
\caption{\label{fig:finite_MI}
${\cal I}$ in the RIM for finite $\ell=2$ for fixed-$h$ ($+$) and box-$h$ ($\Box$) disorders. Upper Inset: Two-point estimates of the $\eta$ exponent are compatible with the analytic SDRG result $\eta=\frac{3-\sqrt{5}}{2}\approx0.382$ (illustrated by the blue line in the main panel). Lower Inset: ${\cal E}$ in the RIM for $\ell=2$.}
\end{figure}

For finite $\ell$, the value of $\cal{B}$ is no longer universal, but asymptotically decays as a power-law with a universal exponent 
\begin{equation}
\mathcal{B}=\frac{c_{\text{eff}}}{3\ln 2}\frac{\ell^2}{r(2\ell+r)}\sim r^{-2}\;,
\end{equation}
which has been checked numerically both with SDRG and DMRG in the RS state 
\cite{calabrese}. In the RIM, ${\cal E}$ is not expected to be universal, as the $p(r)$ probability of a small cluster surviving decimation up to a length scale $~r$ depends heavily on the initial disorder distribution. Two such, independent, surviving clusters 
can connect with a finite probability in the SDRG procedure, resulting in the non-universal expectation ${\cal E}\sim\left(p(r)\right)^2$. In line with this expectation, our numerical results are illustrated in the lower inset of Fig.~\ref{fig:finite_MI}, indicating fast, non-universal decay.

As for ${\cal I}-{\cal E}$, we expect qualitatively different behavior than $\mathcal{B}(r)\sim r^{-2}$ in the RS phase. For $\ell=1$, ${\cal I}-{\cal E}$ is the (longitudinal) two-point correlation function of two spins being in the same cluster. At the critical point, such correlations decay algebraically, with the same exponent for any finite $\ell$, ${\cal I}-{\cal E}\sim r^{-\eta}$ for $r\gg\ell$, where $\eta=\frac{3-\sqrt{5}}{2}\approx0.382$ according to SDRG results \cite{fisher,danielreview}. The value of $\eta$ has been also confirmed by numerical calculations using free-fermionic techniques \cite{young_rieger96,bigpaper}. Our numerical SDRG results for $\ell=2$ are in line with this theoretical expectation (Fig.~\ref{fig:finite_MI}), as both ${\cal I}$ and ${\cal I}-{\cal E}$ are found to decay algebraically with a universal exponent $\eta=0.37(2)$. 

\section{Discussion}
We presented our results for two multipartite entanglement measures ${\cal E}$ and ${\cal I}$ in critical RIM chains between subsystems of size $\ell$, separated at a distance $r$.
Our results indicate a linear scaling limit ($r=\alpha\ell$) for critical quantum correlations, where multipartite entanglement approaches distance-independent universal values for ${\cal E}$ and ${\cal I}$. 
Our results differ qualitatively from those known in critical RS states (e.g. random antiferromagnetic Heisenberg and XX models), where ${\cal I}/{\cal E}=2$. In the RIM, the ratio of ${\cal I}$ and ${\cal E}$ is still universal but depends strongly on $\alpha$, indicating that the two measures capture different universal aspects of the underlying correlations. The strong deviation from the simple RS results is due to the more complex ground state structure in the RIM and is expected to be a generic feature of critical quantum systems that are not in an RS state. We have shown, that the RS results serve as a lower bound for ${\cal I}$ and an upper bound for ${\cal E}$. While we provided quantitative arguments in the Appendix that fit the $\alpha$-dependence of ${\cal E}$ and ${\cal I}$ in the RIM more closely than the RS results, it remains an open challenge to calculate ${\cal E}$ and ${\cal I}$ analytically, similarly to the calculation of the entanglement entropy \cite{refael_moore04}. A potentially promising direction is through the well-known mapping to a random potential representation of the SDRG treatment, similarly to the application in Ref.~\cite{population}.


In addition to the linear scaling limit, there are several other possibilities to consider. A simple case is for adjacent intervals when $r=0$ but $\ell\propto L$. However, this limit does not provide additional information compared to a single subsystem. For both the RIM and models in the RS state, ${\cal E}(\ell)={\cal S}(\ell)/2$, as the same cluster contributions are encountered as for ${\cal S}$, over one endpoint out of the two endpoints of the subsystem. We have also explored the case of finite subsystems (constant $\ell$ when $r\to\infty$), and found a non-universal $\cal E$, while the decay of $\cal I$ is universal and the same as the known SDRG result of the two-point correlation function. In sum, characterizing multipartite entanglement between large \emph{non-adjacent} subsystems is a promising strategy to gain additional insights into quantum criticality compared to single subsystem studies. 
We note that ${\cal I}$ and ${\cal E}$ are expected to behave differently outside of the critical point. While both ${\cal I}$ and ${\cal E}$ have to vanish in the paramagnetic phase, in the ferromagnetic phase only ${\cal E}$ is expected to vanish, while ${\cal I}=1$ for large sizes, as there is a giant magnetic domain.

Our investigations can be extended in several directions.
Here, we mention further quantum chains both with and without disorder, as well as interacting quantum systems in higher dimensions \cite{eisler,EPL,multi}, and systems with long-range interactions \cite{long-range3d,long-rangeCP, long-range}. 
The mutual information (${\cal I}$) results can be readily extended to even more subsystems, $A_1,\dots, A_n$, $n>2$, potentially leading to even more pronounced differences between various models. For $n>2$, ${\cal I}$ is zero in RS states, while non-zero and presumably universal in the RIM. 


\section{Acknowledgments}

We thank Z. Zimbor\'as and R. Juh\'asz 
for helpful discussions. We would like to acknowledge the WCAS Summer Grant Award from the Weinberg College Baker Program in Undergraduate Research at Northwestern University. 



\section{Appendix}
Here, we present two arguments when $r=\alpha L$ as an attempt to better capture ${\cal E}(\alpha)$ and ${\cal I}(\alpha)$ than the $\mathcal{B}$ bound.

\subsection{Estimating ${\cal I}-{\cal E}$}
\begin{figure}[ht!]
\begin{center}
\includegraphics[width=.8\linewidth]{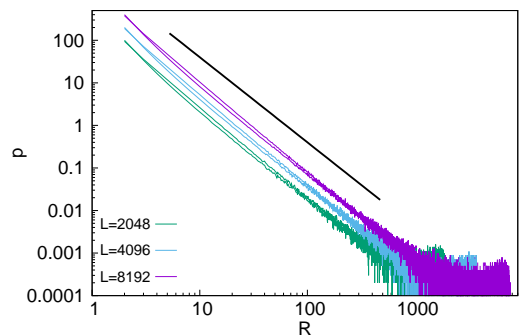}
\end{center}
\vskip-6mm
\caption{\label{fig:app:MI}
Distribution of the linear extent $R$ of the clusters averaged over $10,000$ samples. For each size, we show two lines, corresponding to the box-$h$ and fixed-$h$ distributions, respectively. The black line has a slope of $-2$, illustrating our approximation $p_R\sim LR^{-2}$.}
\end{figure}

${\cal I}-{\cal E}=C_{101}+C_{111}$ has a contribution at most $R-r$ positions out of $L$ for each cluster of linear extent $R>r$, see Fig.~\ref{fig:ill} for an illustration. 
SDRG suggests that there can be only $O(L/R)$ such large clusters. Hence, the position averaged ${\cal I}-{\cal E}$ is expected to be proportional to $1-r/R$ for each such cluster ($R>r$), the probability of which is given by $p_R$. 
Hence,
\begin{multline}
{\cal I}-{\cal E}\propto\smashoperator{\int_{r}^{L}}\mathrm{d}R\,p_R\left(1-\frac{r}{R}\right)
\sim a+b\left(\beta+\frac{1}{\beta}\right),
\label{MI-EN}
\end{multline}
 with $\beta\equiv\frac{r}{L}=\frac{\alpha}{2(2+\alpha)}$. Based on numerical observations (see Fig.~\ref{fig:app:MI}), we used $p_R\sim LR^{-2}$ to estimate the integral. The best fit is given by $a=0.41(3)$ and $b=0.007(2)$.

\subsection{Estimating ${\cal E}$}
Depending on the shape of the cluster (given by $s$, $w_1$ and $w_2$), there are three distinct configurations to consider that contribute to ${\cal E}$. For two equal-sized subsystems of length $\ell$, we can choose to label the cluster such that $w_1 \geq w_2$ without loss of generality. The three contributing cases are: 
\begin{enumerate}
    \item when the range of contributing positions is dictated by $s$. This is the case when $w_1+s\leq\ell+r$ and $w_2+s\leq\ell+r$. There are $s-r$ contributing positions, yielding
    \begin{equation*}
\smashoperator{\int_0^\ell} \mathrm{d}w_1 \smashoperator{\int_0^{w_1}} \mathrm{d}w_2 \smashoperator{\int_r^{\ell+r-w_1}} \mathrm{d}s \,p_{s,w_1,w_2}\frac{s-r}{L}\;,
\end{equation*}
\item when the range of contributing positions is dictated by $s$ on one side and $w_1$ on the other. This is the case when $w_1+s\geq\ell+r$ and $w_2+s\leq\ell+r$. There are $\ell-w_1$ contributing positions, yielding
    \begin{equation*}
\smashoperator{\int_0^\ell} \mathrm{d}w_1 \smashoperator{\int_0^{w_1}} \mathrm{d}w_2 \smashoperator{\int_{\ell+r-w_1}^{\ell+r-w_2}} \mathrm{d}s \,p_{s,w_1,w_2}\frac{\ell-w_1}{L}\;,
\end{equation*}
\item when the range of contributing positions is dictated by $w_1$ on one side and $w_2$ on the other. This is the case when $w_1+s\geq\ell+r$ and $w_2+s\geq\ell+r$ (and $w_1+w_2+s \leq 2\ell+r$). There are $2\ell+r-s-w_1-w_2$ contributing positions, yielding
    \begin{equation*}
\smashoperator{\int_0^\ell}  \mathrm{d}w_1 \smashoperator{\int_0^{w_1}}  \mathrm{d}w_2 \smashoperator{\int_{\ell+r-w_2}^{2\ell+r-w_1-w_2}}  \mathrm{d}s \,p_{s,w_1,w_2}\frac{2\ell+r-s-w_1-w_2}{L}\;.
\end{equation*}
\end{enumerate}
\begin{figure}[ht!]
    \centering
    \includegraphics[width=0.49\textwidth]{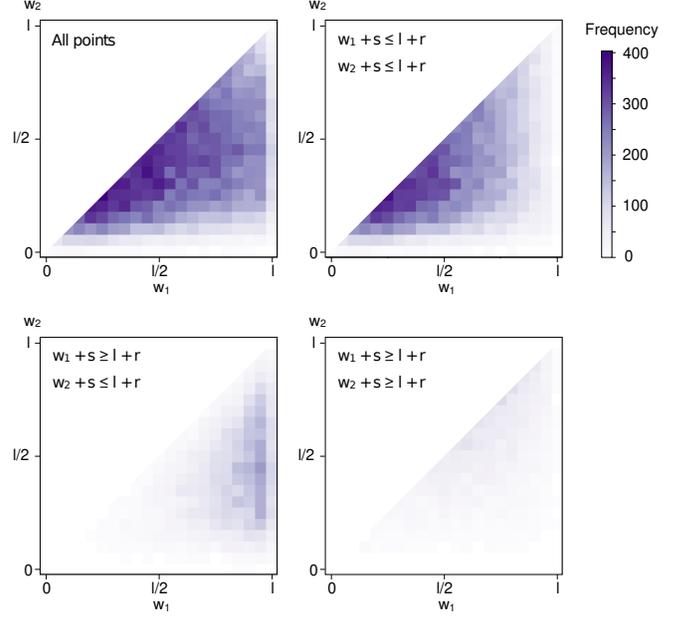}
    \caption{Frequency distribution of clusters contributing to ${\cal E}$ for a system with \(L=8,192\), at \(\alpha\approx 1/2\), which corresponds to \(\ell = 1638.4\) and \(r = 819.2\). The \(w_1\) and \(w_2\) values are binned into bins of width 80. The color of the square representing each bin indicates the number of data points within that bin that contribute to ${\cal E}$. The first panel corresponds to all points that contribute to ${\cal E}$ ($41,966$ points). The other three panels correspond to the points that contribute to each of the three cases discussed and have $28,133$, $10,213$ and $3,620$ points respectively for cases 1, 2, and 3. The color scale is the same for all plots.}
    \label{fig:heatmap}
\end{figure}


The sum of these three contributions is an exact result for ${\cal E}$, assuming that the joint distribution $p_{s,w_1,w_2}$ is known.
As this is not the case, we need to rely on approximations. 
%
As we have cluster contributions only when \(w_1, w_2 \leq \ell\) and \(s\geq r\), along with \(r = \alpha \ell\), we find that the relevant range is \(w_1, w_2 \leq s/\alpha\). 

As a first approximation, we can estimate $w_1=\eta_1s/\alpha$ and $w_2=\eta_2s/\alpha$, yielding 
\begin{multline}
        \frac{c_{\text{eff}}}{3\ln2}\left[\ln\left(\frac{(1+\alpha)(\alpha+\eta_2)}{(\alpha+\eta_1)^2}\right)+\right.\\ \left.\frac{\alpha+\eta_1+\eta_2}{\alpha}\ln\left(\frac{(1+\alpha)(\alpha+\eta_1+\eta_2)}{(2+\alpha)(\alpha+\eta_2)}\right)\right]\;.
\label{eq:EN_app}
\end{multline}
Fitting this to the numerical data, we find \(\eta_1 = 0.87(1)\) and \(\eta_2 = 0.86(1)\), as shown in Fig.~\ref{fig:alpha}. 
While such a fit works in the studied range of $\alpha$, it goes negative for larger $\alpha$. Ideally, we would need an estimate of $p_{s,w_1,w_2}$ that stays positive for any $\alpha$.

As a second approximation, we assume that $w_i$ are independent from each other as well as from $s$, and carry out the $s$ integral first, yielding
\begin{align}
   \smashoperator{\int_0^\ell}  \mathrm{d}w_1 p_{w_1} \smashoperator{\int_0^{w_1}}  \mathrm{d}w_2 p_{w_2}\mathcal{B}(x)\;,
\end{align}
where $x$ depends on $w_1$ and $w_2$, through $\ell_1=\ell-w_1$ and $\ell_2=\ell-w_2$ in Eq.~(\ref{eq:x}). This result is intuitive as it indicates that additional sites in the cluster reduce the number of contributing positions by effectively reducing the size of the subsystems by the amount $w_i$.
Assuming a box distribution for both $w_1$ and $w_2$ leads to
\begin{multline}
\label{eq:EN_app2}
    \frac{1}{(2+\alpha)^2}\left[2 (1+\alpha) (3+\alpha) \ln \left(\frac{1+\alpha }{\alpha }\right)- \right.\\
    \left.(2+\alpha)^2 \ln \left(\frac{2+\alpha}{\alpha }\right)-1\right]\;,
\end{multline}
up to 
an overall multiplicative constant. In Fig.~\ref{fig:alpha} we show Eq.~(\ref{eq:EN_app2}) fitted to the data, with the constant numerically determined. While the fit is less good than the first approximation over the range of \(\alpha\) values investigated, Eq.~(\ref{eq:EN_app2}) remains positive in the limit \(\alpha\to\infty\), scaling as \(1/\alpha^4\). 


\end{document}